\documentclass[aps, twocolumn, prd, preprintnumbers, amsmath, amssymb, amsfonts, superscriptaddress,  nofootinbib]{revtex4-1}

\pdfoutput=1

\usepackage[linktocpage,pagebackref=false,hidelinks]{hyperref}
\usepackage[pagebackref=false,hidelinks]{hyperref}
\usepackage{diagbox}
\usepackage{latexsym}
\usepackage{color}
\usepackage{epsfig}
\usepackage{graphicx}
\usepackage[export]{adjustbox}
\usepackage[dvipsnames]{xcolor}
\usepackage{float}
\usepackage{tabularx}
\usepackage{placeins}
\usepackage{multirow}

\newcommand{\beq}{\begin{equation}}
\newcommand{\eeq}{\end{equation}}
\newcommand{\bea}{\begin{eqnarray}}
\newcommand{\eea}{\end{eqnarray}}
\newcommand{\nn}{\nonumber}

\newcommand{\eV}{\mathrm{eV}}
\newcommand{\keV}{\mathrm{keV}}
\newcommand{\GeV}{\mathrm{GeV}}
\newcommand{\MeV}{\mathrm{MeV}}
\newcommand{\TeV}{\mathrm{TeV}}

\def\bal#1\eal{\begin{align}#1\end{align}}

\newcommand{\vev}[1]{\left\langle #1\right\rangle}
\newcommand{\Lag}{\mathcal{L}}
\newcommand{\order}{\mathcal{O}}

\begin{document}

\title{Environment-dependent mass splitting to suppress solar capture in the inelastic-doublet interpretation of the LZ event}

\author{Sudhakantha Girmohanta}
\email{sudhakantha5@gmail.com}
\affiliation{Particle Theory  and Cosmology Group (PTC),
Center for Theoretical Physics of the Universe (CTPU), \\
Institute for Basic Science, Daejeon 34126, Republic of Korea}

\author{Tae Hyun Jung}
\email{thjung0720@gmail.com}
\affiliation{Particle Theory  and Cosmology Group (PTC),
Center for Theoretical Physics of the Universe (CTPU), \\
Institute for Basic Science, Daejeon 34126, Republic of Korea}

\begin{abstract}
The inelastic electroweak-doublet interpretation of the $248\,{\rm keV}$ recoil energy event reported by LUX-ZEPLIN faces strong constraints from solar capture. We propose an environment-dependent mass splitting generated by an ultralight scalar coupled quadratically to electrons. Its density-induced expectation value enhances the splitting in the Sun while preserving the laboratory value, providing a mechanism to suppress solar capture. We derive conditions for this mechanism and examine constraints from stellar cooling and electron-mass variations. 
We find that this scenario is phenomenologically viable, while it suffers from severe fine-tuning of the scalar potential.
\end{abstract}

\maketitle

\section{Introduction}
\label{sec:introduction}

The LUX-ZEPLIN (LZ) Collaboration has recently extended its dark-matter
search to nuclear-recoil energies of approximately $270 \,\keV$ using
an exposure of $2.84$ tonne-years~\cite{LZ:2026axp}.  
One event was observed at
\begin{equation}
E_R=248\pm23\,({\rm stat})\pm23({\rm sys})\,\keV \, ,
\label{eq:LZevent}
\end{equation}
in a region with a very low expected background.  
The largest local significance among the signal hypotheses considered by LZ is $3.4\sigma$, which is reduced to a global significance of $2.6\sigma$ after accounting for the look-elsewhere effect. 
Although this event is not statistically significant evidence for dark matter, its unusually large recoil energy motivates particle-physics interpretations with a kinematic threshold.

A simple possibility is an inelastic dark matter scenario with a vector-like pair of pseudo-Dirac electroweak doublets with a large Dirac mass while two neutral states are split by a small Majorana mass term induced after electroweak symmetry breaking.
Their mass eigenstates possess approximately maximal mixing of opposite hypercharge states, and therefore satisfy
\begin{equation}
\chi_1+{\cal N}\longrightarrow\chi_2+{\cal N} \, ,
\label{eq:endothermic}
\end{equation}
where $\chi_1$ and $\chi_2$ are the light and heavy states, respectively, and ${\cal N}$ is the target nucleus.
Interestingly, this interpretation can be realized in a supersymmetric setup with the Higgsino dark matter\,\cite{Fan:2026kxx, Langhoff:2026ujr, Yin:2026jnn, Wu:2026nhi, Freese:2026sga, DiMauro:2026ldr, Du:2026guj, Bisal:2026khf, Khan:2026zuj}, and also in a non-supersymmetric setup\,\cite{ Visinelli:2026kgt, Kotlarski:2026pep, Ahmed:2026qjg, Paul:2026okh}.

Since it is mediated by the $Z$ boson, the dark matter-nucleus cross section is governed by its typical neutron scattering cross section $\sigma_{\chi_1 n} \simeq 7.3\times 10^{-39}\,{\rm cm}^{2}$ and the atomic and mass numbers of the target nucleus\,\cite{Goodman:1984dc, Essig:2007az, Pospelov:2026ewn, Rodd:2026tyn}.
With a benchmark mass of $1.1 \,\mathrm{TeV}$ motivated by the relic abundance from the thermal freeze-out scenario, the LZ signal event can be explained by the mass splitting $\delta\sim 350 \,\keV$ for the Standard Halo Model and $\sim 490\,\keV$ for a halo model incorporating Large Magellanic Cloud effects\,\cite{Fan:2026kxx, Langhoff:2026ujr}\footnote{
However, a null signal event in the high-energy sideband disfavors the pure doublet dark matter interpretation\,\cite{Rodd:2026tyn, Dent:2026bji, Delepine:2026ith}.
The strength of the tension depends on the signal acceptance in the high recoil energy bin, which must be investigated further.
We do not discuss this tension in this work.
}.
If the doublets are mixed additionally with a singlet component, the nucleon cross section can be lowered, and a smaller dark matter mass can explain the signal\,\cite{Borah:2026zwf, Lee:2026jxl, Nagata:2026pbj}.
In this work, we only consider the pure doublet case, although there is a large room for various interpretations and their constraints\,\cite{Nomura:2026qyq, Wang:2026ytg, Bandyopadhyay:2026gjw, Elahi:2026vlm, Yamashita:2026ump, Jeesun:2026vzo, Unwin:2026rdp, Smirnov:2026aqk, deLima:2026shq, Lee:2026wof, Das:2026uyy, Liang:2026coz, Yang:2026wpb, Kannike:2026qyl, Du:2026lpa, Alhazmi:2026efz, Okada:2026eol, Lee:2026xxh, Asadi:2026iot, Aghaie:2026vsu, Zhu:2026dag, Yuan:2026djt, Kumar:2026lgi, Qi:2026vyp, He:2026hqz, Heikinheimo:2026kwp, Fan:2026hzw, Su:2026rwz, McCabe:2026crm, Gu:2026vto, Khan:2026nwp, Chattaraj:2026fxn, Okada:2026upm, Mahapatra:2026glu, DiMauro:2026ymt, He:2026idw, Uttayarat:2026isp, Palmisano:2026kuj, Baer:2026yrt, Ahmed:2026kan, Lian:2026hpm, Das:2026buc, Le-Yaouanc:2026djt, Barman:2026omh, Borah:2026ris, Xing:2026civ, Ge:2026xax, An:2026pkc, Arcadi:2026kev, Okada:2026fef, Bose:2026szs, OHare:2026nqi, De:2026win, Ahmed:2026com, Khan:2026osp, Sheng:2026tqt, Chauhan:2026udz, Sannino:2026hkc, Gemmell:2026yaw, Jung:2026otm}.

The same interaction of \eqref{eq:endothermic} also captures doublet dark matter in the Sun. 
Then, the captured dark matter population annihilates into a pair of weak gauge bosons, leaving a high-energy neutrino signal to be tested at Earth.
The absence of such a neutrino signal in the IceCube neutrino observatory implies that the mass splitting must be larger than approximately $570\,\keV$\,\cite{Pospelov:2026ewn, DiMauro:2026dqp, Bose:2026ndd, Nguyen:2026lui, Ghosh:2026txe}.
It appears to exclude the inelastic pure doublet explanation of the LZ signal event.

This conclusion, however, can be changed if $\delta$ at Earth and the Sun are different, which is the case investigated in this work.
In particular, we consider the possibility that the mass splitting depends on the expectation value of a scalar field whose effective potential is sensitive to the ambient matter density.  
The scalar expectation value vanishes in the laboratory but becomes nonzero above a critical density, enhancing $\delta$ in the solar core.  
The goal is to obtain the mass splitting at the Sun greater than $570\,\keV$ while keeping it around $350\,\keV$ at Earth.

We present our mechanism as a low-energy effective theory.
Taking the electron density as a representative environmental source, we derive the conditions for the density-induced scalar background and examine the leading constraints from the environmental electron-mass shift and stellar cooling.
Similar ideas of environment-dependent dark matter or dark sector interactions have also been studied in other contexts~\cite{Khoury:2003aq, Masso:2005ym, Masso:2006gc, Jaeckel:2006xm, Kim:2007wj, Brax:2007ak, Redondo:2007lda, Hinterbichler:2010es, Bloch:2020uzh, DeRocco:2020xdt, Chakraborty:2020vec}.

\section{The mechanism}
\label{sec:mechanism}

\subsection{Doublet dark matter}
We consider a vector-like pair of doublet Weyl fermions
\begin{equation}
 L=\begin{pmatrix}N\\ E\end{pmatrix} \, ,
 \qquad
 L^c=\begin{pmatrix}E^c\\ N^c\end{pmatrix} \, ,
 \label{eq:doublets}
\end{equation}
whose hypercharges are $-1/2$ and $1/2$, respectively.
A Dirac mass connects $L$ and $L^c$, and the Majorana masses of $N$ and $N^c$ are generated after electroweak (EW) symmetry breaking, while it depends on the expectation value of a complex scalar field $\Phi$.
The effective Lagrangian after EW symmetry breaking is then written as
\begin{align}
 -\Lag_{\rm DM}& \! \supset \! M_D N N^c \!
 +\! \frac{1}{2} m_{M}(\phi) N N \!+\! \frac{1}{2} m_{M}'(\phi) N^c N^c \!+\! {\rm h.c.},
 \label{eq:LDM}
\end{align}
where the Majorana masses $m_M$ and $m_M'$ are functions of expectation values of $\phi$, the radial component of $\Phi$.
The stability of the dark matter can be protected by imposing a $Z_2$ symmetry, under which $L$ and $L^c$ are odd, and all Standard Model fields and $\Phi$ are even.
In the supersymmetric setup, one can identify $L$ and $L^c$ as down- and up-type Higgsinos, and the $Z_2$ symmetry as $R$-parity.
$E$ and $E^c$ are approximately $350\,\MeV$ heavier than $N$ and $N^c$ for $M_D \sim \TeV$ due to electroweak corrections, and their disappearing track signal can be tested at future colliders\,\cite{Cheung:2026byg}.

For $m_M,\,m_M'\ll M_D$, the two mass eigenstates $\chi_1$ and $\chi_2$
have masses
\begin{equation}
 m_{1,2}=M_D\mp\frac{m_M+m_M'}{2}
 +\order\!\left(\frac{m_M^2}{M_D},\,\frac{m_M'^2}{M_D}\right) \, ,
\end{equation}
denoting $\chi_1$ ($\chi_2$) as the light (heavy) state.  
Here, we assume that $M_D$, $m_M$ and $m_M'$ are all real for simplicity.
Their mass splitting is
\begin{equation}
 \delta\equiv m_{\chi_2}-m_{\chi_1}
 =m_M + m_M' \, ,
 \label{eq:splittingphi}
\end{equation}
and the $Z$ current interaction becomes off-diagonal at leading order up to $|m_M-m_M'|/M_D$. 
Therefore, the dominant tree-level nuclear scattering is nearly the transition in Eq.~\eqref{eq:endothermic}. 

In order to explain the LZ signal, we take $M_D$ and $\delta$ at the LZ laboratory environment
\bal
M_D \simeq \TeV, \quad \delta =\delta_{\rm LZ}\simeq 350\,\keV
\quad \text{(at LZ)} \, ,
\eal
assuming the Standard Halo Model\,\cite{Fan:2026kxx}.
To avoid the solar capture bound, we need\,\cite{Pospelov:2026ewn, DiMauro:2026dqp, Bose:2026ndd, Nguyen:2026lui, Ghosh:2026txe}
\bal
\delta \gtrsim \delta_\odot \simeq 570\,\keV \quad \text{(at the sun)} \, ,
\eal
which seemingly rules out the doublet explanation of the LZ signal event.

We avoid this tension by a density-dependent expectation value of $\phi$, making $m_M$ and $m_M'$ enhanced in the solar core environment.
For this, we introduce effective interactions
\bal
\frac{\phi^2}{2\Lambda_N} N N + \frac{\phi^2}{2\Lambda_{N^c}} N^c N^c + {\rm h.c.}\,  
\label{eq:Yukawa}
\eal
with nonzero bare Majorana masses $\frac{1}{2}m_{M0} N N + \frac{1}{2}m_{M0}' N^c N^c + {\rm h.c.}$,
while making the $\phi$ expectation value $v_\phi$ density-dependent. 
Here, the quadratic coupling is chosen because a linear coupling would generate a dangerous tadpole with a nonzero $m_{M0}$ and $m_{M0}'$.
Such a linear coupling can be forbidden by imposing a $U(1)$ global symmetry under which $\Phi$ is charged.
Now, the question is how to make $\phi$ density-dependent.

For simplicity, let us set $m_{M0}=m_{M0}'$ and $\Lambda_N=\Lambda_{N^c}$.
Since $\delta \ll M_D$, the elastic scattering induced by the hierarchy between $m_M$ and $m_M'$ can be neglected, and thus introducing a hierarchically different $m_M$ and $m_M'$ does not change the discussion below.

\subsection{Density-dependent scalar background}

We assume that the quadratic term in the $\phi$ potential is sensitive to the background density.
In particular, we take the form of 
\begin{align}
 V(\phi)=&\frac{\lambda_\phi}{4} \phi^4 + \Big(\frac{\mu_\phi^2}{2} -\frac{m_e}{2\Lambda_e^2} \, \bar e  e\Big) \phi^2 \, ,
 \label{eq:Vphi}
\end{align}
where the expectation of the $\bar e  e$ operator becomes
\begin{equation}
 \vev{\bar e e}
 \simeq n_e
\end{equation}
in a non-relativistic electron medium with its number density $n_e$.
The negative sign of the coefficient in the $\bar e e \phi^2$ operator indicates that a nonzero scalar expectation value can be developed for a large electron density.

We take $\mu_\phi^2 >0$, so $v_\phi = 0$ around Earth, to avoid strong constraints from equivalence principle tests; if $\phi$ gets a nonzero vev in the laboratory environment, the effective Yukawa coupling to electrons is generated as $y_{\rm eff} = v_\phi m_e/\Lambda_e^2$ with $v_\phi \equiv \langle \phi \rangle$, whose constraint is as strong as $y_{\rm eff} \lesssim 10^{-25}$ for a light $\phi$\,\cite{Antypas:2022asj}.

Therefore, we have a critical electron density above which $v_\phi$ becomes nonzero,
\bal
v_\phi^2(n_e) = 
\begin{cases}
    \frac{m_e n_e - \mu_\phi^2\Lambda_e^2}{\lambda_\phi \Lambda_e^2}
    &
    \text{for $n_e > n_e^c \equiv \frac{\mu_\phi^2\Lambda_e^2}{m_e}$} \, ,
    \\
    0
    &
    \text{for $n_e < n_e^c$} \, .
\end{cases}
\eal
In order to have the effect at the Sun, $n_e^c$ has to be smaller than the solar core density $n_e^\odot \simeq 500\,\keV^3$.
This condition leads to an upper bound on $\Lambda_e$
\bal
\Lambda_e & < \frac{\sqrt{m_e n_e^\odot}}{\mu_\phi}
\simeq  10^{11}\,\GeV 
\bigg( \frac{4 \times 10^{-12}\,\eV}{\mu_\phi} \bigg) \, .
\label{eq:Lambda_e_upper}
\eal
In addition, the Compton wavelength of $\phi$, namely $ \mu_\phi^{-1}$, has to be shorter than the size of the solar core $R_{\odot, {\rm core}} \sim (10^{-15}\,\eV)^{-1}$.
Therefore, we have
\bal
\mu_\phi \gtrsim 10^{-15}\,\eV \, .
\label{eq:condition2}
\eal
Otherwise, $\phi$ would see the density averaged over its Compton-wavelength volume, which reduces the effective electron number density significantly from $n_e^\odot$.
We do not explore a parameter region in such a case.

In order to maintain $v_\phi=0$ around Earth, we need the critical density to be greater than the electron density at Earth.
Since the Earth radius is $R_\oplus\sim (3\times 10^{-14}\,\eV)^{-1}$, the Compton wavelength of $\phi$ can be greater than the Earth radius.
Therefore, we estimate the effective electron number density around Earth to be the one averaged over the Compton-wavelength volume
\bal
\bar n_e^\oplus \sim 10\,\keV^3 \times \min \bigg[ \bigg( \frac{\mu_\phi}{3\times 10^{-14}\,\eV} \bigg)^{\!\! 3}, \, 1 \bigg].
\label{Eq:comptonAvg}
\eal
From $n_e^c > \bar n_e^\oplus$, we have a lower bound on $\Lambda_e$ as
\bal
\Lambda_e &> 2\times 10^{12}\,\GeV \nn\\
&\quad 
\times \min \bigg[ \bigg( \frac{\mu_\phi}{3\times 10^{-14}\,\eV} \bigg)^{1/2}, \, \bigg(\frac{3\times 10^{-14}\,\eV}{\mu_\phi} \bigg) \bigg].
\label{eq:Lambda_e_lower}
\eal
This, together with Eq.\,\eqref{eq:Lambda_e_upper}, implies that the window allowed for $\Lambda_e$ is not large.
For simplicity, we fix $\Lambda_e$ to the geometric mean of its maximum and minimum allowed values, obtained from Eqs.~\eqref{eq:Lambda_e_upper} and \eqref{eq:Lambda_e_lower}, respectively, and denote it by $\bar{\Lambda}_e(\mu_\phi)$ for each $\mu_\phi$ in our parameter scan.

The mass splitting now becomes a function of $n_e$
\bal
\delta(n_e) = 2\left(m_{M0} + \frac{v_\phi^2(n_e)}{\Lambda_N}\right)  .
\eal
In the laboratory, since $v_\phi=0$, the constant Majorana mass has to be $m_{M0} = \delta_{\rm LZ}/2$ to explain the LZ signal event.

At the Sun, we need the mass splitting greater than $\delta_\odot$ to avoid the solar capture bound.
Defining $\Delta \delta$ to be the acquired mass splitting in the Sun, we have
\bal
\Delta \delta &\equiv \delta(n_e^\odot) - \delta(\bar n_e^\oplus)
=
\frac{2}{\Lambda_N}\,\frac{m_e n_e^\odot - \mu_\phi^2 \Lambda_e^2}{\lambda_\phi\Lambda_e^2} 
\\
&\simeq 250\,\keV \bigg( \frac{10^{-50}}{\lambda_\phi} \bigg) \! \bigg(\frac{7\times 10^{13}\,\GeV}{\Lambda_N} \bigg) \! \bigg(\frac{5 \times 10^{10}\,\GeV}{\Lambda_e} \bigg)^{\!\! 2},
\label{eq:delta_solar_capture}
\eal
with $\Delta \delta > \delta_\odot - \delta_{\rm LZ} \simeq 220\,\keV$.
Once this and Eqs.\,\eqref{eq:Lambda_e_lower}, \eqref{eq:condition2} and \eqref{eq:Lambda_e_upper} are satisfied, the LZ experiment can be explained by pure doublet inelastic dark matter without the solar capture bound.

\begin{figure}[t]
    \centering
    \includegraphics[width=0.49\textwidth]{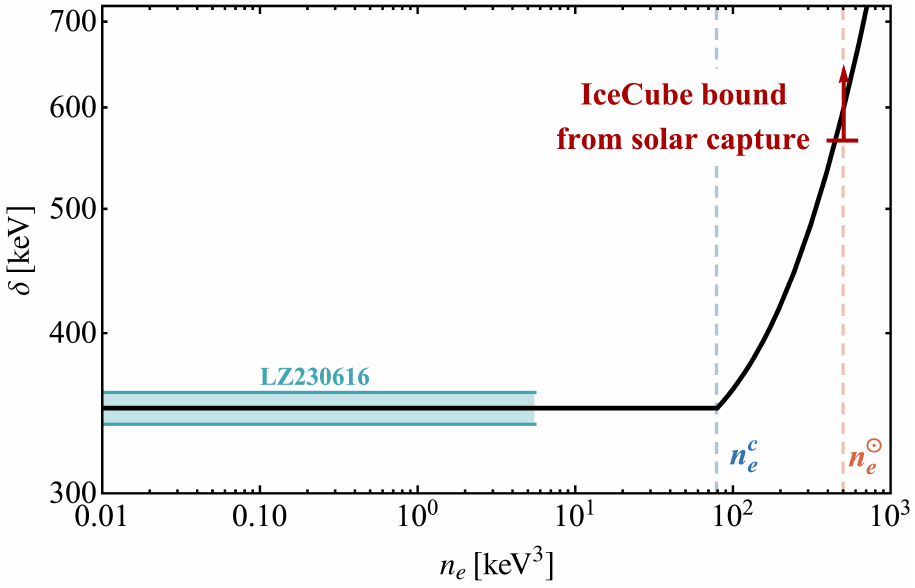}
    \caption{{Mass splitting $\delta$ as a function of the local electron density
$n_e$. Below the critical density $n_e^c$, the splitting is fixed at $2m_{M0}=350 \, \keV$, consistent with the LZ signal (cyan band), whereas it increases with $n_e$ for $n_e>n_e^c$. The benchmark parameters are taken as $\mu_\phi=4 \times 10^{-12} \, \eV$, and $\Lambda_N=7 \times 10^{13}~\GeV$. We fix $\Lambda_e$ to the geometric mean of its maximum and minimum allowed values for the given $\mu_\phi$ from Eqs.~\eqref{eq:Lambda_e_upper}, and~\eqref{eq:Lambda_e_lower}, namely $\Lambda_e = \bar{\Lambda}_e \simeq 5 \times 10^{10} \, \GeV$. This yields $n_e^c\simeq 79 \, \keV^3$, indicated by the blue dashed line. At the solar electron density (red dashed line), the splitting reaches $\delta(n_e^\odot)=600 \, \keV$, thereby evading the IceCube bound on solar capture. The corresponding lower limit on the splitting is shown by the red solid line and arrow. $\lambda_\phi$ is fixed at $\simeq 10^{-50}$ corresponding to this chosen benchmark through Eq.~\eqref{eq:delta_solar_capture}.}}
\label{fig:Delta_m12}
\end{figure}

In Fig.\,\ref{fig:Delta_m12}, we present $\delta$ as a function of $n_e$.
Parameters are chosen to explain the LZ signal event (indicated by the cyan band) while avoiding the solar capture bound (indicated by the red arrow). We choose the representative parameter values $m_{M0}= \delta_{\rm LZ}/2=175~\keV$, $\Lambda_N= 7 \times 10^{13} \, \GeV$ and $\mu_\phi=4 \times 10^{-12} \, \eV$ (which leads to $\bar{\Lambda}_e \simeq 5 \times 10^{10} \, \GeV$).
For this parameter choice, one obtains $n_e^c\simeq 79~\keV^3$. For concreteness,
we set $\delta(n_e^\odot)=600~\keV$, which yields $\lambda_\phi \simeq 10^{-50}$ following Eq.~\eqref{eq:delta_solar_capture}. The red dashed vertical line corresponds to the electron density in the solar core, while the blue dashed one denotes $n_e^c$.

The flatness of the $\phi$ potential requires a severe fine-tuning of the effective potential whose quantum corrections are generated by $L$, $L^c$, and the electron loops.
For instance, we find that the level of tuning $\lambda_\phi$ is $O(10^{-26})$ for $\Lambda_N\sim10^{16}\,\GeV$ and $\Lambda_e\sim10^{11}\,\GeV$.
This is essentially because the one-loop diagrams of them are quadratically divergent, and the scale of $M_D$ is too large compared to the characteristic scales $n_e \simeq 500\,\keV^3$, $\Delta \delta \simeq 200\,\keV$, and $m_e \simeq 0.5\,\MeV$.
Slightly milder, but still severe, tuning is also required for higher-dimensional operators $\phi^6$ and $\phi^8$.
We checked different but similar setups (e.g., linear coupling of $\phi$ to $N N$ and $N^c N^c$, the $\mu_\phi^2<0$ case and/or the $m_{M0}=0$ case, etc.), but they all required a similar or higher level of fine-tuning.
This fine-tuning may indicate that our mechanism is unnatural.
Nevertheless, we accept it, just as we accept the fine-tuning of the quadratic term in the SM Higgs potential.

\section{Phenomenological constraints}
\label{sec:constraints}

\begin{table}[t]
\begin{tabular}{c|ccc}
    & $n_e$ at core & core size & $y_{\rm eff}^{\rm upper}$ \\ 
    \hline \hline
Sun~ & ~$500\,\keV^3$~   &  ~$(10^{-15}\,\eV)^{-1}$~ &  $2.4\times 10^{-13}$\,\cite{Gondolo:2008dd} \\
HB~  & ~$2\times10^4\,\keV^3$~  &  ~$(3\times 10^{-15}\,\eV)^{-1}$~ & ~$3\times 10^{-15}$\,\cite{Hardy:2016kme}~   \\
RG~  & ~$2\times10^6\,\keV^3$~ & ~$(2\times 10^{-14}\,\eV)^{-1}$~  & ~$7\times 10^{-16}$\,\cite{Hardy:2016kme} \\
WD~  & ~$4 \times 10^6 \, \keV^3$~ & ~$(3\times 10^{-14}\,\eV)^{-1}$~ & ~$4\times 10^{-16}$\,\cite{Bottaro:2023gep} 
\end{tabular}
\caption{Summary of scales of electron density and radius for various stellar cores, and corresponding upper bound on $y_{\rm eff}$.}
\label{Tab:stellar_bounds}
\end{table}

Although an effective electron Yukawa with $\phi$ is not generated around Earth,  it can be generated in the core of stars such as the Sun, red giants (RGs), horizontal branch (HB) stars, white dwarfs (WDs), etc.
The effective electron Yukawa coupling can be written as
\bal
&y_{\rm eff}(n_e) = \frac{m_e v_\phi(n_e)}{\Lambda_e^2}
\\ \nonumber
&\quad \sim 2 \times 10^{-20}
\bigg( \frac{n_e}{n_e^\odot}\bigg)^{1/2}
\bigg( \frac{\Delta \delta}{250\,\keV} \bigg)^{1/2} \\
&  \qquad \qquad \times
\bigg( \frac{\Lambda_N}{7 \times 10^{13}\,\GeV} \bigg)^{1/2}
\bigg( \frac{ 5 \times 10^{10}\,\GeV}{\Lambda_e} \bigg)^2 \, .
\label{eq:y_eff}
\eal
There are upper bounds on $y_{\rm eff}$ from the stellar cooling argument for the Sun\,\cite{Gondolo:2008dd}\footnote{
Searching for $\phi$ emitted from the Sun provides a stronger bound on $y_{\rm eff}$ for the usual case\,\cite{Budnik:2019olh}, but it does not apply in our case, since $v_\phi =0$ at Earth.
}, HB stars, RG\,\cite{Hardy:2016kme}, and WD~\cite{Bottaro:2023gep}.
We summarize them in the last column of Table\,\ref{Tab:stellar_bounds}, denoting the upper bound on $y_{\rm eff}$ as $y^{\rm upper}_{\rm eff}$. 

Since $y_{\rm eff} \propto n_e^{1/2}$, a larger-density star tends to provide a stronger bound.
Indeed, we find that the most stringent constraint comes from the considerations of the WD luminosity function of the galactic disk, for which $y_{\rm eff} \lesssim 4 \times 10^{-16}$~\cite{Bottaro:2023gep}.

The interaction in Eq.\,\eqref{eq:Vphi} makes the electron mass density-dependent. In our scenario, the fractional electron mass shift is given as follows
\bal
&\frac{|\Delta m_e|}{m_e} = \frac{v_\phi^2(n_e)}{{2}\Lambda_e^2} \simeq 
\frac{n_e}{n_e^\odot} \frac{\Delta \delta \, \Lambda_N}{{4}\Lambda_e^2}
\\ \nonumber
& \quad \simeq 2 \times 10^{-8} 
\bigg( \frac{n_e}{4 \times 10^6 \, \keV^3}\bigg)
\bigg( \frac{\Delta \delta}{250\,\keV} \bigg) \\
& \qquad \qquad \qquad \times
\bigg( \frac{\Lambda_N}{7 \times 10^{13}\,\GeV} \bigg) 
 \bigg( \frac{5 \times10^{10} \,\GeV}{\Lambda_e} \bigg)^2,
\eal
which is small, but can be relevant for dense stars, especially for smaller $\Lambda_e$ and larger $\Lambda_N$. Among the various stellar objects, WD provides the most stringent constraint. In particular, the mass and radius measurement for Sirius B is in excellent agreement with the theoretical mass–radius relation for a WD~\cite{Bond_2017}. A change in electron mass results in a change in the WD radius as 
\begin{align}
    \frac{\delta R_{\rm WD}}{R_{\rm WD}} = -\frac{\Delta m_e}{m_e} \ .
    \label{Eq:RWDchange}
\end{align}
With the measured precision of the Sirius B radius, we adopt $|\Delta m_e/m_e|_{\rm Sirius \, B} \lesssim 1\%$ as a criterion to be consistent with observation. We use $n_e \simeq  10^{7} \, \keV^3$ for Sirius B, and its radius as $(3.5 \times 10^{-14} \, \eV)^{-1}$.

Furthermore, since $v_\phi$ grows with $n_e$, we also consider early-Universe constraints from the induced electron mass shift. For $T \gg m_e$ the thermal condensate is $\langle\bar e e\rangle_T \simeq m_e T^2/6$, resulting in
\begin{align}
  v_\phi^2(T) \simeq \frac{m_e^2 T^2}{6\,\lambda_\phi \Lambda_e^2}\, .
  \label{Eq:veffT}
\end{align}
The resulting shift in the electron mass can alter primordial light element abundances in Big Bang nucleosynthesis (BBN) at the time of neutron-proton freeze-out, namely around $T_n \sim \MeV$.
Based on the study\,\cite{Garramone:2026evc}, we adopt a criterion
\begin{align}
& \left|\frac{\Delta m_e}{m_e}\right|_{T_n} \simeq 0.07\frac{m_e^2 T_n^2}{\lambda_\phi \Lambda_e^4} \lesssim 1\% \, ,
  \label{Eq:meBBN}
  \end{align}
where we have taken the numerical integration and obtained $\langle\bar e e \rangle_{T_n} \simeq 0.14 m_e T_n^2$.

\begin{figure}[t]
    \centering
    \includegraphics[width=0.99\linewidth]{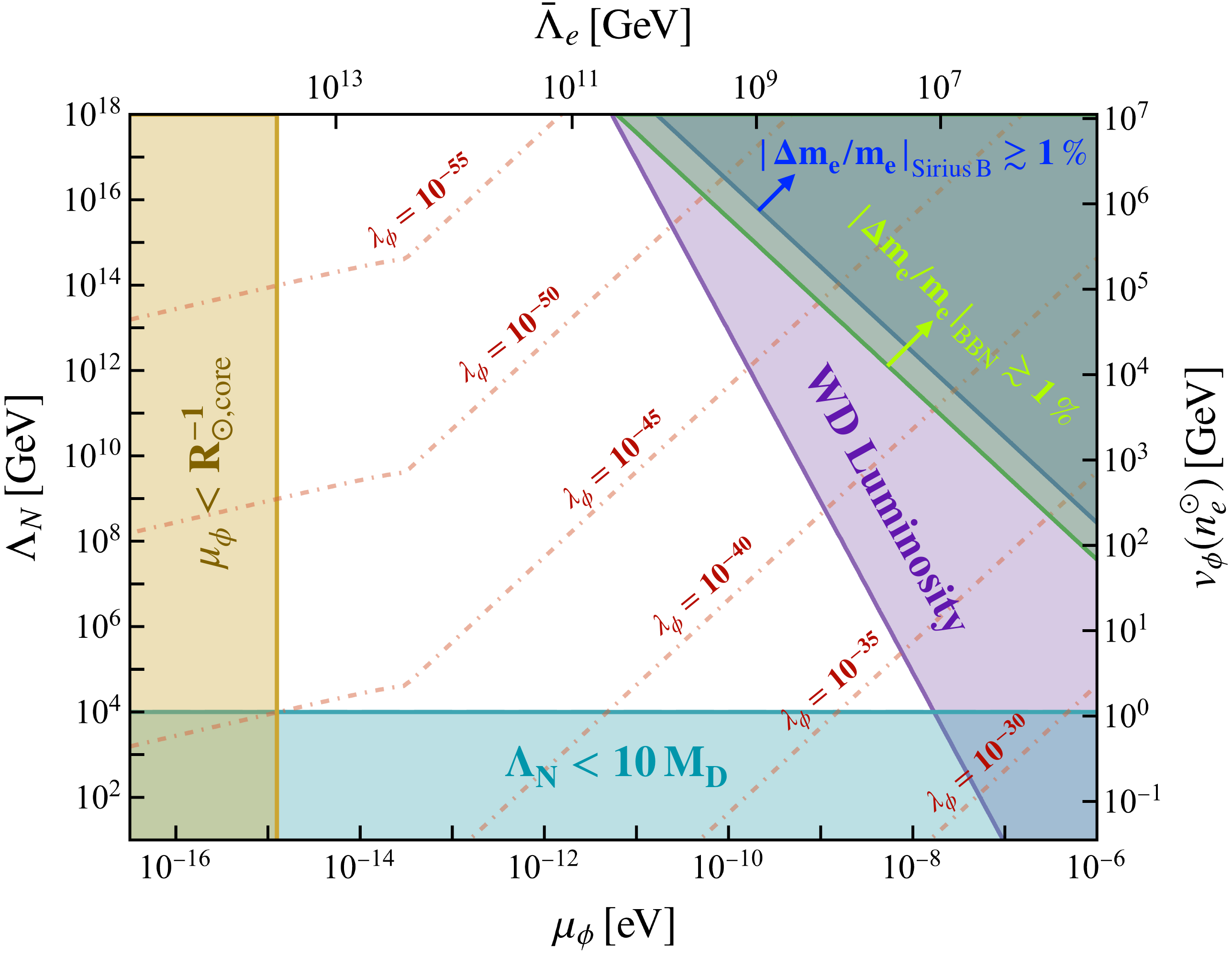}
    \caption{{Constraints in the $\mu_\phi$--$\Lambda_N$ plane for $\delta(n_e^\odot) = 600 \, \keV$, which suffices to evade the IceCube bound from solar capture. Here $\Lambda_e$ is fixed to $\bar{\Lambda}_e(\mu_\phi)$, the geometric mean of the maximum and minimum values of $\Lambda_e$ allowed at a given $\mu_\phi$ by Eqs.~\eqref{eq:Lambda_e_upper} and~\eqref{eq:Lambda_e_lower}. The brown region is unexplored in this work due to the finite solar size condition, $\mu_\phi < R_{\odot,\rm core}^{-1}$. The dominant constraints over the allowed mass range come from the WD luminosity function of the galactic disk (purple)~\cite{Bottaro:2023gep}, from the electron mass shift in Sirius B (blue)~\cite{Bond_2017}, and during BBN (green)~\cite{Garramone:2026evc}. The cyan region is excluded by EFT validity, which we impose as $\Lambda_N > 10\,M_D$. Red dashed contours give the corresponding values of $\lambda_\phi$. The white region is phenomenologically viable. The associated $v_\phi(n_e^\odot)$ and $\bar{\Lambda}_e$ are shown on the right and top axes, respectively.}}
    \label{fig:results}
\end{figure}

In Fig.~\ref{fig:results} we show the relevant phenomenological constraints discussed above in the $\mu_\phi$--$\Lambda_N$
plane, fixing $\Lambda_e = \bar{\Lambda}_e(\mu_\phi)$ at each $\mu_\phi$ and determining
$\lambda_\phi$ from the chosen mass splitting in the Sun, $\delta(n_e^\odot) = 600~\keV$,
which suffices to evade the IceCube bound from solar capture. 
We only explore $\mu_\phi$ that follows from the finite-size condition \eqref{eq:condition2}, so our analysis does not apply to the brown region.
Within the allowed range, the dominant constraints come from WD cooling (purple) and
from the electron-mass shift during BBN (green) and in Sirius B (blue). The cyan
region is a lower bound on $\Lambda_N$ from the validity of the EFT treatment, which we take
as $\Lambda_N > 10\,M_D$. The white region is phenomenologically viable. Contours of
$\lambda_\phi$ are overlaid as red dashed lines, and the corresponding values of
$v_\phi(n_e^\odot)$ and $\bar{\Lambda}_e$ appear on the right and top axes, respectively.

\section{Discussion}
\label{sec:conclusion}

We have presented an environment-dependent resolution of the tension between the inelastic-doublet interpretation of the LZ high-energy recoil event and the solar-capture constraint. 
The LZ signal favors a mass splitting $350 \, \keV$, whereas suppressing solar capture requires a splitting above approximately $570 \, \keV$.  
These requirements are not inconsistent if the Majorana masses, and hence $\delta$, depend on the local matter density, especially in the Sun.

The essential ingredient is a scalar whose effective mass squared changes sign above a critical density.  
The scalar expectation value vanishes in the laboratory but becomes nonzero in the solar interior. 
Consequently, the laboratory splitting is determined by the density-independent Majorana mass, $\delta=2m_{M0}$ at the LZ experiment, while the solar splitting receives the additional contribution $2v_\phi^2(n_e^\odot)/\Lambda_N$.
This separates the two relevant kinematic thresholds responsible for the LZ scattering event and solar capture.
While our scenario is phenomenologically viable as shown in Fig.\,\ref{fig:results}, the major obstacle is the theoretical requirement of fine-tuning on the effective potential of $\phi$.

Although we have described our scenario in the low-energy effective theory, let us outline its possible UV completion.
The effective operator \eqref{eq:Yukawa} requires a UV model that connects $\Phi$ and the EW symmetry-breaking sector, since $N$ and $N^c$ belong to EW doublets.
The required interaction can arise from gauge-invariant operators such as $|\Phi|^2 (HL)^2$ with the SM Higgs doublet $H$ (or an additional doublet $D$ with a small vev).
Such operators can originate from renormalizable interactions involving additional singlet or triplet fermions in analogy with the type-I and III seesaw models.
Alternatively, one can consider a hypercharge $+1$ triplet scalar $\Delta$ with a small vev.
The interactions $L \Delta L$ and $|\Phi|^2 |\Delta|^2$ then allow the Majorana masses to depend on $|\Phi|^2$ through the triplet vev.
The electron coupling in Eq.\,\eqref{eq:Vphi} can be obtained by a positive mixed quartic term $|H|^2 |\Phi|^2$ (or $|D|^2|\Phi|^2$) in the scalar potential, while the Higgs portal will also introduce the nucleon density dependence, from which we expect a similar scenario and phenomenology to apply.
These possibilities provide starting points for UV realizations and variants of our scenario. 
A detailed study of their phenomenology and thermal histories is left for future work.

\begin{acknowledgments}
We thank Hyung Do Kim and Kohsaku Tobioka for useful discussion. This work was supported by IBS under the project code IBS-R018-D1.
\end{acknowledgments}

\bibliography{references}

@article{LZ:2026axp,
    author = "Akerib, D. S. and others",
    collaboration = "LZ",
    title = "{Search for dark matter particle interactions in an extended nuclear recoil energy window with the LUX-ZEPLIN (LZ) experiment}",
    eprint = "2609.02823",
    archivePrefix = "arXiv",
    primaryClass = "hep-ex",
    doi = "10.17182/hepdata.182472.v1",
    month = "9",
    year = "2026"
}

@article{Yin:2026jnn,
    author = "Yin, Wen",
    title = "{A PQ-Symmetric High-Scale SUSY Interpretation of the LZ High-Energy Recoil}",
    eprint = "2609.01892",
    archivePrefix = "arXiv",
    primaryClass = "hep-ph",
    month = "9",
    year = "2026"
}

@article{Fan:2026kxx,
    author = "Fan, JiJi and Reece, Matthew",
    title = "{Higgsino Above the Sea of Fog}",
    eprint = "2609.01504",
    archivePrefix = "arXiv",
    primaryClass = "hep-ph",
    month = "9",
    year = "2026"
}

@article{Wu:2026nhi,
    author = "Wu, Lei and Zhang, Yang and Zhu, Bin",
    title = "{TeV Higgsino Dark Matter from LZ Nuclear Recoil to Fermi-LAT Gamma Rays}",
    eprint = "2609.01590",
    archivePrefix = "arXiv",
    primaryClass = "hep-ph",
    month = "9",
    year = "2026"
}

@article{Freese:2026sga,
    author = "Freese, Katherine and Theodosopoulos, Dionysios P.",
    title = "{Higgsino Dark Matter Interpretation of the LUX-ZEPLIN 248 keV Nuclear-Recoil Event}",
    eprint = "2609.01583",
    archivePrefix = "arXiv",
    primaryClass = "hep-ph",
    month = "9",
    year = "2026"
}

@article{Su:2026rwz,
    author = "Su, Liangliang and Yang, Jin Min and Yang, Wen-Na",
    title = "{Inelastic Dark Matter Signature at High Recoil Energy in LUX-ZEPLIN and CRESST}",
    eprint = "2609.01475",
    archivePrefix = "arXiv",
    primaryClass = "hep-ph",
    month = "9",
    year = "2026"
}

@article{Yamashita:2026ump,
    author = "Yamashita, Kimiko",
    title = "{Inelastic Dark Photon Dark Matter for the LUX-ZEPLIN High-Recoil Event and the Galactic Halo Gamma-Ray Excess}",
    eprint = "2609.02868",
    archivePrefix = "arXiv",
    primaryClass = "hep-ph",
    month = "9",
    year = "2026"
}

@article{Nomura:2026qyq,
    author = "Nomura, Yasunori",
    title = "{Dark Matter as the {$Z_2$} Partner of the Standard Model Higgs Boson}",
    eprint = "2609.02505",
    archivePrefix = "arXiv",
    primaryClass = "hep-ph",
    reportNumber = "RIKEN-iTHEMS-Report-26",
    month = "9",
    year = "2026"
}

@article{Visinelli:2026kgt,
    author = "Visinelli, Luca",
    title = "{A Peccei-Quinn Origin for Inelastic Electroweak Dark Matter after LUX-ZEPLIN}",
    eprint = "2609.02807",
    archivePrefix = "arXiv",
    primaryClass = "hep-ph",
    month = "9",
    year = "2026"
}

@article{Pospelov:2026ewn,
    author = "Pospelov, Maxim and Ramani, Harikrishnan",
    title = "{Strong Constraints on Higgsino Dark Matter from Solar Capture}",
    eprint = "2609.02775",
    archivePrefix = "arXiv",
    primaryClass = "hep-ph",
    month = "9",
    year = "2026"
}

@article{DiMauro:2026ldr,
    author = "Di Mauro, Mattia",
    title = "{Dark Matter at the Kinematic Edge: Interpreting the 248 keV LZ Nuclear-Recoil Candidate}",
    eprint = "2609.02608",
    archivePrefix = "arXiv",
    primaryClass = "hep-ph",
    month = "9",
    year = "2026"
}

@article{Rodd:2026tyn,
    author = "Rodd, Nicholas L. and Safdi, Benjamin R. and Slatyer, Tracy R. and Xu, Weishuang Linda",
    title = "{Confronting the Higgsino Interpretation of the LZ Event with the High-Energy Sideband}",
    eprint = "2609.04175",
    archivePrefix = "arXiv",
    primaryClass = "hep-ph",
    month = "9",
    year = "2026"
}

@article{Jeesun:2026vzo,
    author = "Jeesun, Sk and Majumdar, Anirban",
    title = "{Atmospheric neutrino up-scattering explanation of LZ 2026 excess}",
    eprint = "2609.04185",
    archivePrefix = "arXiv",
    primaryClass = "hep-ph",
    month = "9",
    year = "2026"
}

@article{McCabe:2026crm,
    author = "McCabe, Christopher",
    title = "{Seasonal dark matter from the LUX-ZEPLIN high-energy event}",
    eprint = "2609.04181",
    archivePrefix = "arXiv",
    primaryClass = "hep-ph",
    month = "9",
    year = "2026"
}

@article{Du:2026guj,
    author = "Du, Xiaokang and Wang, Fei",
    title = "{TeV Higgsino Interpretation of the LZ High-Recoil Event with Intermediate-Scale Electroweak Gauginos}",
    eprint = "2609.04163",
    archivePrefix = "arXiv",
    primaryClass = "hep-ph",
    month = "9",
    year = "2026"
}

@article{Unwin:2026rdp,
    author = "Unwin, James",
    title = "{Axion Portal Dark Matter and the LUX-ZEPLIN High-Recoil Event}",
    eprint = "2609.04186",
    archivePrefix = "arXiv",
    primaryClass = "hep-ph",
    month = "9",
    year = "2026"
}

@article{Smirnov:2026aqk,
    author = "Smirnov, Juri and Griffith, Spencer and Beacom, John F.",
    title = "{Inelastic Signatures of Electroweak Dark Matter}",
    eprint = "2609.04144",
    archivePrefix = "arXiv",
    primaryClass = "hep-ph",
    month = "9",
    year = "2026"
}

@article{deLima:2026shq,
    author = "de Lima, Carlos Henrique",
    title = "{Exothermic Dark Matter at LZ}",
    eprint = "2609.05204",
    archivePrefix = "arXiv",
    primaryClass = "hep-ph",
    month = "9",
    year = "2026"
}

@article{Gu:2026vto,
    author = "Gu, Guanhua and Li, Lingfeng and Tang, Shao-Song and Xu, Yongheng",
    title = "{Inelastic from the Other Side: Xenon Excitation Signals in Light of the LZ High-Recoil Event}",
    eprint = "2609.05291",
    archivePrefix = "arXiv",
    primaryClass = "hep-ph",
    month = "9",
    year = "2026"
}

@article{Dent:2026bji,
    author = "Dent, James B. and Newstead, Jayden L.",
    title = "{Exothermic and Endothermic Inelastic Dark Matter Interpretations at LZ: Sideband Constraints and Future Prospects}",
    eprint = "2609.04673",
    archivePrefix = "arXiv",
    primaryClass = "hep-ph",
    month = "9",
    year = "2026"
}

@article{Lee:2026wof,
    author = "Lee, Hyun Min",
    title = "{Inelastic dark matter and baryon flavor symmetry in light of LUX-ZEPLIN (LZ) experiment}",
    eprint = "2609.06171",
    archivePrefix = "arXiv",
    primaryClass = "hep-ph",
    month = "9",
    year = "2026"
}

@article{Das:2026uyy,
    author = "Das, Pritam and Karmakar, Biswajit and Mahapatra, Satyabrata and Paul, Partha Kumar",
    title = "{Inelastic Self-interacting Dark Matter and LUX-ZEPLIN 248 keV Event in a Dirac Modular Inverse Seesaw}",
    eprint = "2609.06825",
    archivePrefix = "arXiv",
    primaryClass = "hep-ph",
    month = "9",
    year = "2026"
}

@article{Kotlarski:2026pep,
    author = "Kotlarski, Wojciech and Kowalska, Kamila and Sessolo, Enrico Maria",
    title = "{GUT-induced FCC signatures of the LUX-ZEPLIN event}",
    eprint = "2609.06750",
    archivePrefix = "arXiv",
    primaryClass = "hep-ph",
    month = "9",
    year = "2026"
}

@article{Liang:2026coz,
    author = "Liang, Jin-Han and Liu, Zuowei and Tran, Van Que and Xu, Yongheng",
    title = "{LZ Nuclear-Recoil Excess from Boosted Light Magnetic Dipole-dipole Dark Matter}",
    eprint = "2609.06756",
    archivePrefix = "arXiv",
    primaryClass = "hep-ph",
    month = "9",
    year = "2026"
}

@article{DiMauro:2026dqp,
    author = "Di Mauro, Mattia and Shaikh, Halim",
    title = "{Solar Capture Tests of Inelastic Dark Matter after the LZ High-Recoil Event}",
    eprint = "2609.06760",
    archivePrefix = "arXiv",
    primaryClass = "hep-ph",
    month = "9",
    year = "2026"
}

@article{Yang:2026wpb,
    author = "Yang, Meiwen and Wu, Quan-feng and Tsai, Yue-Lin Sming and Fan, Yi-Zhong",
    title = "{Multi-Messenger and Paleo-Detector Probes of the LZ Dark Matter Signal}",
    eprint = "2609.06640",
    archivePrefix = "arXiv",
    primaryClass = "hep-ph",
    month = "9",
    year = "2026"
}

@article{Wang:2026ytg,
    author = "Wang, Lei and Xiao, Yang",
    title = "{The Inert Doublet Model of Dark Matter and the LUX-ZEPLIN High-Recoil Event}",
    eprint = "2609.06571",
    archivePrefix = "arXiv",
    primaryClass = "hep-ph",
    month = "9",
    year = "2026"
}

@article{Khan:2026nwp,
    author = "Khan, Imtiaz and Capozziello, Salvatore and Mustafa, G. and Atamurotov, Farruh and Abdujabbarov, Ahmadjon and Yuan, Chengxun",
    title = "{Nuclear interference versus dark sector excitation in the 248 keV LUX-ZEPLIN recoil candidate}",
    eprint = "2609.09230",
    archivePrefix = "arXiv",
    primaryClass = "hep-ph",
    month = "9",
    year = "2026"
}

@article{Bose:2026ndd,
    author = "Bose, Debajit and others",
    title = "{Not so good {$\nu$}s for Higgsino dark matter as LZ excess: stringent limits from Super-Kamiokande and IceCube}",
    eprint = "2609.07807",
    archivePrefix = "arXiv",
    primaryClass = "hep-ph",
    month = "9",
    year = "2026"
}

@article{Kannike:2026qyl,
    author = "Kannike, Kristjan and Raidal, Martti and Strumia, Alessandro",
    title = "{Boosted dark particles and the LZ nuclear recoil event}",
    eprint = "2609.07742",
    archivePrefix = "arXiv",
    primaryClass = "hep-ph",
    month = "9",
    year = "2026"
}

@article{Bisal:2026khf,
    author = "Bisal, Subhadip and Cao, Junjie and Li, Fei",
    title = "{Higgsino Dark Matter Interpretation of the LZ High-Recoil Event in the GNMSSM with TeV-Scale Gauginos}",
    eprint = "2609.07811",
    archivePrefix = "arXiv",
    primaryClass = "hep-ph",
    month = "9",
    year = "2026"
}

@article{Borah:2026zwf,
    author = "Borah, Debasish and Sahoo, Sujit Kumar and Sahu, Narendra and Sharma, Shashwat",
    title = "{Inelastic Singlet-Doublet Fermion Dark Matter in light of the 248 keV LZ event}",
    eprint = "2609.07800",
    archivePrefix = "arXiv",
    primaryClass = "hep-ph",
    month = "9",
    year = "2026"
}

@article{Bandyopadhyay:2026gjw,
    author = "Bandyopadhyay, Disha and Borah, Debasish and Borah, Pankaj",
    title = "{LZ nuclear recoil event from inelastic singlet-doublet scalar dark matter}",
    eprint = "2609.07451",
    archivePrefix = "arXiv",
    primaryClass = "hep-ph",
    month = "9",
    year = "2026"
}

@article{Du:2026lpa,
    author = "Du, Xin-Yu and Huang, Wenjie and Xie, Keping",
    title = "{Pseudo-Dirac Inelastic Dark Matter in the Leptophobic $U(1)_B$ Model: Confronting the LUX-ZEPLIN High-Recoil Event with Collider Searches}",
    eprint = "2609.07225",
    archivePrefix = "arXiv",
    primaryClass = "hep-ph",
    month = "9",
    year = "2026"
}

@article{Ahmed:2026qjg,
    author = "Ahmed, Waqas and Leontaris, George K.",
    title = "{A Dark-Dimension Origin of Geometric Inelastic Dark Matter: The LUX-ZEPLIN High-Recoil Event and Multi-Target Tests}",
    eprint = "2609.07138",
    archivePrefix = "arXiv",
    primaryClass = "hep-ph",
    month = "9",
    year = "2026"
}

@article{Alhazmi:2026efz,
    author = "Alhazmi, Haider and Kim, Doojin and Kong, Kyoungchul and Park, Jong-Chul and Shin, Seodong",
    title = "{High-Energy Nuclear Recoils from Boosted Dark Matter for the LZ 248-keV Event: Beyond the Halo-Dependent High-Velocity Tail}",
    eprint = "2609.06890",
    archivePrefix = "arXiv",
    primaryClass = "hep-ph",
    month = "9",
    year = "2026"
}

@article{Okada:2026eol,
    author = "Okada, Nobuchika and Seto, Osamu",
    title = "{Inelastic $B-L$ scalar dark matter and the LUX-ZEPLIN event}",
    eprint = "2609.06909",
    archivePrefix = "arXiv",
    primaryClass = "hep-ph",
    reportNumber = "EPHOU-26-011",
    month = "9",
    year = "2026"
}

@article{Langhoff:2026ujr,
    author = "Langhoff, Kevin",
    title = "{Heavy Higgsino Interpretation of the LZ Event}",
    eprint = "2609.09385",
    archivePrefix = "arXiv",
    primaryClass = "hep-ph",
    month = "9",
    year = "2026"
}

@article{Lee:2026xxh,
    author = "Lee, Vincent S. H. and Randall, Lisa",
    title = "{A Warped Extra Dimensional Candidate for the LZ 248 keV Event}",
    eprint = "2609.09136",
    archivePrefix = "arXiv",
    primaryClass = "hep-ph",
    reportNumber = "N3AS-26-020",
    month = "9",
    year = "2026"
}

@article{Asadi:2026iot,
    author = "Asadi, Pouya and Batz, Austin and Fox, Patrick J. and Homiller, Samuel D. and Kribs, Graham D.",
    title = "{For Whom the Xenon Recoils: Magnetic Inelastic Dark Baryons}",
    eprint = "2609.09107",
    archivePrefix = "arXiv",
    primaryClass = "hep-ph",
    reportNumber = "PITT-PACC-2614, FERMILAB-PUB-26-0660-T",
    month = "9",
    year = "2026"
}

@article{Lee:2026jxl,
    author = "Lee, Seung J. and Youn, Taewook",
    title = "{Mixing-suppressed inelastic dark matter: a minimal model for the LZ 248 keV event}",
    eprint = "2609.09138",
    archivePrefix = "arXiv",
    primaryClass = "hep-ph",
    month = "9",
    year = "2026"
}

@article{Aghaie:2026vsu,
    author = "Aghaie, Mohammad and Strumia, Alessandro",
    title = "{Neutron disappearance and the LZ nuclear recoil event}",
    eprint = "2609.09037",
    archivePrefix = "arXiv",
    primaryClass = "hep-ph",
    month = "9",
    year = "2026"
}

@article{Zhu:2026dag,
    author = "Zhu, Pengxuan and Dalla Valle Garcia, Giovani and Wang, Xuan-Gong and Thomas, Anthony W. and White, Martin J.",
    title = "{Endothermic dark matter with a light dark photon and the LUX--ZEPLIN high-energy nuclear-recoil candidate}",
    eprint = "2609.09015",
    archivePrefix = "arXiv",
    primaryClass = "hep-ph",
    month = "9",
    year = "2026"
}

@article{Elahi:2026vlm,
    author = "Elahi, Fatemeh and Schwaller, Pedro",
    title = "{A Vector-Like Lepton Interpretation of the High-Energy Nuclear Recoil Candidate in LUX-ZEPLIN}",
    eprint = "2609.08993",
    archivePrefix = "arXiv",
    primaryClass = "hep-ph",
    reportNumber = "MITP-26-043",
    month = "9",
    year = "2026"
}

@article{Cheung:2026byg,
    author = "Cheung, Kingman and Kang, Sin Kyu and Kumar, Ranjeet",
    title = "{From LUX-ZEPLIN to Colliders: Probing Higgsino Dark Matter}",
    eprint = "2609.08712",
    archivePrefix = "arXiv",
    primaryClass = "hep-ph",
    month = "9",
    year = "2026"
}

@article{Yuan:2026djt,
    author = "Yuan, Guan-Wen and Zhang, Bo and Cao, Wen-Yu and Feng, Lei and Yang, Ruizhi",
    title = "{ALP-mediated inelastic dark matter and the LUX-ZEPLIN high-recoil candidate event LZ230616}",
    eprint = "2609.08893",
    archivePrefix = "arXiv",
    primaryClass = "hep-ph",
    month = "9",
    year = "2026"
}

@article{Kumar:2026lgi,
    author = "Kumar, Ranjeet and Prajapati, Hemant Kumar",
    title = "{Generalized Chiral $U(1)_{B-L}$ with Inelastic Scalar Dark Matter for the LZ 248 keV Event}",
    eprint = "2609.10827",
    archivePrefix = "arXiv",
    primaryClass = "hep-ph",
    month = "9",
    year = "2026"
}

@article{Qi:2026vyp,
    author = "Qi, XinXin and Sun, Hao",
    title = "{Solar Capture and Suppressed Annihilation of Inelastic Scalar Dark Matter}",
    eprint = "2609.10636",
    archivePrefix = "arXiv",
    primaryClass = "hep-ph",
    month = "9",
    year = "2026"
}

@article{He:2026hqz,
    author = "He, Yuxuan",
    title = "{Transition magnetic-dipole dark matter and the LZ230616 high-recoil candidate}",
    eprint = "2609.10453",
    archivePrefix = "arXiv",
    primaryClass = "hep-ph",
    month = "9",
    year = "2026"
}

@article{Chattaraj:2026fxn,
    author = "Chattaraj, Ayan and Majumdar, Anirban and Papoulias, Dimitrios K. and Srivastava, Rahul",
    title = "{Can Elastic Neutrino Scattering Account for the LZ230616 Event?}",
    eprint = "2609.10504",
    archivePrefix = "arXiv",
    primaryClass = "hep-ph",
    month = "9",
    year = "2026"
}

@article{Fan:2026hzw,
    author = "Fan, Zi-Tong and He, Hong-Jian and Wang, Yu-Chen and Zhao, Yue",
    title = "{Inelastic Dark Matter and High-Energy Recoil Signatures in LZ}",
    eprint = "2609.10491",
    archivePrefix = "arXiv",
    primaryClass = "hep-ph",
    month = "9",
    year = "2026"
}

@article{Nguyen:2026lui,
    author = "Nguyen, Thong T. Q. and Linden, Tim and Hooper, Dan",
    title = "{Solar Neutrino Constraints on Inelastic Dark Matter Scattering in Light of Recent LUX-ZEPLIN Observations}",
    eprint = "2609.11833",
    archivePrefix = "arXiv",
    primaryClass = "hep-ph",
    month = "9",
    year = "2026"
}

@article{Heikinheimo:2026kwp,
    author = "Heikinheimo, Matti and Zimmermann, Niklas",
    title = "{Cosmic ray boosted dark matter with momentum dependent interactions can explain the LZ 248 keV event}",
    eprint = "2609.11600",
    archivePrefix = "arXiv",
    primaryClass = "hep-ph",
    month = "9",
    year = "2026"
}

@article{Okada:2026upm,
    author = "Okada, Hiroshi and Shigekami, Yoshihiro and Wu, Jia-Jun",
    title = "{Can a minimal radiative seesaw explain the LZ 248 keV event?}",
    eprint = "2609.13038",
    archivePrefix = "arXiv",
    primaryClass = "hep-ph",
    month = "9",
    year = "2026"
}

@article{Mahapatra:2026glu,
    author = "Mahapatra, Satyabrata and Paul, Partha Kumar",
    title = "{Boosted or Inelastic? Discriminating Interpretations of the LZ 248 keV Event}",
    eprint = "2609.14799",
    archivePrefix = "arXiv",
    primaryClass = "hep-ph",
    month = "9",
    year = "2026"
}

@article{DiMauro:2026ymt,
    author = "Di Mauro, Mattia",
    title = "{Testing Higgs-Coupled Minimal Dark Matter with Solar Neutrinos after the LZ High-Recoil Event}",
    eprint = "2609.19174",
    archivePrefix = "arXiv",
    primaryClass = "hep-ph",
    month = "9",
    year = "2026"
}

@article{He:2026idw,
    author = "He, Xiao-Gang and Hong, Xuan and Jeesun, Sk",
    title = "{Hadrophilic inelastic freeze-in dark matter in $q_1-q_2$ gauge extension and the high energy LZ event}",
    eprint = "2609.15714",
    archivePrefix = "arXiv",
    primaryClass = "hep-ph",
    month = "9",
    year = "2026"
}

@article{Uttayarat:2026isp,
    author = "Uttayarat, P. and Julio, J. and Primulando, R.",
    title = "{DM induced neutron disappearance as the origin of the LZ nuclear recoil event}",
    eprint = "2609.15933",
    archivePrefix = "arXiv",
    primaryClass = "hep-ph",
    month = "9",
    year = "2026"
}

@article{Palmisano:2026kuj,
    author = "Palmisano, Stefano and Tammaro, Michele and Tesi, Andrea",
    title = "{Inferring dark matter masses and interactions from high recoil energy events in LUX-ZEPLIN}",
    eprint = "2609.15985",
    archivePrefix = "arXiv",
    primaryClass = "hep-ph",
    month = "9",
    year = "2026"
}

@article{Baer:2026yrt,
    author = "Baer, Howard and Barger, Vernon",
    title = "{Argon as the test of two interpretations of the LZ 248 keV recoil}",
    eprint = "2609.15782",
    archivePrefix = "arXiv",
    primaryClass = "hep-ph",
    month = "9",
    year = "2026"
}

@article{Ahmed:2026kan,
    author = "Ahmed, Waqas and Ahmad, Ammara and Rehman, Mansoor Ur",
    title = "{Xenon Isotope Filtering at the Kinematic Edge of Inelastic Dark Matter}",
    eprint = "2609.15634",
    archivePrefix = "arXiv",
    primaryClass = "hep-ph",
    month = "9",
    year = "2026"
}

@article{Lian:2026hpm,
    author = "Lian, Jingwei and Yang, Jin Min",
    title = "{Explain the LZ High-Energy Recoil Event with Inelastic Sneutrino Dark Matter in Supersymmetry}",
    eprint = "2609.15742",
    archivePrefix = "arXiv",
    primaryClass = "hep-ph",
    month = "9",
    year = "2026"
}

@article{Das:2026buc,
    author = "Das, Arindam and Nomura, Takaaki",
    title = "{Effect of inelastic scalar dark matter in hidden $U(1)$ scenario after the LZ nuclear recoil}",
    eprint = "2609.15600",
    archivePrefix = "arXiv",
    primaryClass = "hep-ph",
    month = "9",
    year = "2026"
}

@article{Ghosh:2026txe,
    author = "Ghosh, Aditya and Chavez, Ilumi and Kelso, Chris",
    title = "{Confronting the Higgsino Interpretation of the LZ Event with Astrophysical Uncertainties and the Solar Capture Constraints}",
    eprint = "2609.15321",
    archivePrefix = "arXiv",
    primaryClass = "hep-ph",
    month = "9",
    year = "2026"
}

@inproceedings{Le-Yaouanc:2026djt,
    author = "Le-Yaouanc, Alain and Richard, Fran{\c{c}}ois",
    title = "{A DM candidate indicated at Fermi-LAT and LZ ? Connection with LHC and LC prospects}",
    booktitle = "{The 2026 International Workshop on Future Linear Colliders}",
    eprint = "2609.15413",
    archivePrefix = "arXiv",
    primaryClass = "hep-ph",
    month = "9",
    year = "2026"
}

@article{Barman:2026omh,
    author = "Barman, Basabendu",
    title = "{Did LZ see modified gravity?}",
    eprint = "2609.15118",
    archivePrefix = "arXiv",
    primaryClass = "hep-ph",
    month = "9",
    year = "2026"
}

@article{Borah:2026ris,
    author = "Borah, Pankaj and Mahapatra, Satyabrata and Nath, Newton",
    title = "{Inelastic Dark Matter at LZ from Radiative Dirac Neutrino Mass Paradigm}",
    eprint = "2609.15027",
    archivePrefix = "arXiv",
    primaryClass = "hep-ph",
    month = "9",
    year = "2026"
}

@article{Xing:2026civ,
    author = "Xing, Chuan-Yang",
    title = "{Galactic Endothermic Production and Exothermic Detection of Excited Dark Matter: Implications for LUX-ZEPLIN}",
    eprint = "2609.17935",
    archivePrefix = "arXiv",
    primaryClass = "hep-ph",
    month = "9",
    year = "2026"
}

@article{Ge:2026xax,
    author = "Ge, Shao-Feng and Titov, Oleg and Wang, Yakun",
    title = "{Dark Matter Inelastic Scattering with Nuclei for Direct Detection}",
    eprint = "2609.16529",
    archivePrefix = "arXiv",
    primaryClass = "hep-ph",
    month = "9",
    year = "2026"
}

@article{An:2026pkc,
    author = "An, Haipeng and Gao, Fei and Liu, Jia and Liu, Minghao and Xu, Changlong",
    title = "{Cosmological Constrained Axion-Portal Inelastic Dark Matter for the LZ Event}",
    eprint = "2609.17412",
    archivePrefix = "arXiv",
    primaryClass = "hep-ph",
    month = "9",
    year = "2026"
}

@article{Arcadi:2026kev,
    author = "Arcadi, Giorgio and di Mauro, Mattia and Djouadi, Abdelhak and Queiroz, Farinaldo",
    title = "{A possible interpretation of the LUX-ZEPLIN recoil event in the 2HD+a scenario}",
    eprint = "2609.17196",
    archivePrefix = "arXiv",
    primaryClass = "hep-ph",
    month = "9",
    year = "2026"
}

@article{Nagata:2026pbj,
    author = "Nagata, Natsumi and Yanagida, Tsutomu T.",
    title = "{Asymmetric Inelastic Dark Matter and the LUX-ZEPLIN event}",
    eprint = "2609.18564",
    archivePrefix = "arXiv",
    primaryClass = "hep-ph",
    month = "9",
    year = "2026"
}

@article{Okada:2026fef,
    author = "Okada, Nobuchika and Raut, Digesh",
    title = "{Endothermic Z'-Portal Dark Matter: LZ-LHC Complementarity}",
    eprint = "2609.21011",
    archivePrefix = "arXiv",
    primaryClass = "hep-ph",
    month = "9",
    year = "2026"
}

@article{Paul:2026okh,
    author = "Paul, Partha Kumar and Sahoo, Sujit Kumar and Sahu, Narendra and Sharma, Shashwat",
    title = "{Resurrecting Electroweak Dark Matter via Type-II Seesaw in light of recent LZ Event}",
    eprint = "2609.22063",
    archivePrefix = "arXiv",
    primaryClass = "hep-ph",
    month = "9",
    year = "2026"
}

@article{Bose:2026szs,
    author = "Bose, Debajit and Saha, Akash Kumar and Raj, Nirmal and Maity, Tarak Nath and Laha, Ranjan",
    title = "{LUX-ZEPLIN's Stairway to Hea{$\nu$}en: limits on elastic scatters of dark matter from solar capture}",
    eprint = "2609.21823",
    archivePrefix = "arXiv",
    primaryClass = "hep-ph",
    month = "9",
    year = "2026"
}

@article{OHare:2026nqi,
    author = "O'Hare, Ciaran A. J.",
    title = "{The high-velocity dark matter halo of the Milky Way in light of the LZ 248 keV event}",
    eprint = "2609.21444",
    archivePrefix = "arXiv",
    primaryClass = "hep-ph",
    month = "9",
    year = "2026"
}

@article{De:2026win,
    author = "De, Bibhabasu",
    title = "{The 248 keV LZ Recoil: A Possible Hint of Non-SM-Like Quark Yukawa Couplings with a Scalar-Portal Dark Matter}",
    eprint = "2609.23096",
    archivePrefix = "arXiv",
    primaryClass = "hep-ph",
    month = "9",
    year = "2026"
}

@article{Ahmed:2026com,
    author = "Ahmed, Waqas and Leontaris, George K.",
    title = "{Kaluza--Klein Dark-Photon Mediation of Inelastic Dark Matter at LUX-ZEPLIN}",
    eprint = "2609.22739",
    archivePrefix = "arXiv",
    primaryClass = "hep-ph",
    month = "9",
    year = "2026"
}

@article{Khan:2026osp,
    author = "Khan, Imtiaz and Capozziello, Salvatore and Mustafa, G. and Botirov, Farkhod and Abdujabbarov, Ahmadjon and Atamurotov, Farruh and Channuie, Phongpichit",
    title = "{Elastic toroidal vector dark matter through a dark photon in the LUX-ZEPLIN high recoil window}",
    eprint = "2609.25114",
    archivePrefix = "arXiv",
    primaryClass = "hep-ph",
    month = "9",
    year = "2026"
}

@article{Khan:2026zuj,
    author = "Khan, Imtiaz and Muhammad, Ali and Mustafa, G. and Atamurotov, Farruh and Abdujabbarov, Ahmadjon and Khan, Mussawir",
    title = "{LZ-Motivated Pseudo-Dirac Higgsinos in the Supersymmetric 331 Model from the Supersymmetric {SU(6)} GUT Model}",
    eprint = "2609.23691",
    archivePrefix = "arXiv",
    primaryClass = "hep-ph",
    month = "9",
    year = "2026"
}

@article{Sheng:2026tqt,
    author = "Sheng, Jie and Zhang, Kairui",
    title = "{The LUX-ZEPLIN Event as Hyperfine Spectroscopy of Composite Dark Matter}",
    eprint = "2609.23477",
    archivePrefix = "arXiv",
    primaryClass = "hep-ph",
    month = "9",
    year = "2026"
}

@article{Chauhan:2026udz,
    author = "Chauhan, Bhavesh and Sahasrabuddhe, Soham and Sen, Manibrata",
    title = "{A closer look at the LZ 248 keV event through the lens of cosmic-ray boosted dark matter}",
    eprint = "2609.24982",
    archivePrefix = "arXiv",
    primaryClass = "hep-ph",
    month = "9",
    year = "2026"
}

@article{Sannino:2026hkc,
    author = "Sannino, Francesco and Turner, Jessica",
    title = "{Interpreting the LZ 248 keV Event using Dark QCD}",
    eprint = "2609.24988",
    archivePrefix = "arXiv",
    primaryClass = "hep-ph",
    month = "9",
    year = "2026"
}

@article{Delepine:2026ith,
    author = "Delepine, David and Khalil, Shaaban",
    title = "{Model-Independent Sideband Constraints on Inelastic Dark Matter at the LZ High-Recoil Candidate}",
    eprint = "2609.26698",
    archivePrefix = "arXiv",
    primaryClass = "hep-ph",
    month = "9",
    year = "2026"
}

@article{Gemmell:2026yaw,
    author = "Gemmell, Caleb and Hooper, Dan and Krnjaic, Gordan",
    title = "{A Simple Dark Matter Model to Explain the LZ Event and Galactic Center Excess}",
    eprint = "2609.26570",
    archivePrefix = "arXiv",
    primaryClass = "hep-ph",
    reportNumber = "FERMILAB-PUB-26-0702-T",
    month = "9",
    year = "2026"
}

@article{Jung:2026otm,
    author = "Jung, Hyunjoo and Park, Seong Chan",
    title = "{Dark Diffraction at LZ from a Screened Neutral Composite Baryon}",
    eprint = "2609.25723",
    archivePrefix = "arXiv",
    primaryClass = "hep-ph",
    month = "9",
    year = "2026"
}

@article{Khoury:2003aq,
    author = "Khoury, Justin and Weltman, Amanda",
    title = "{Chameleon fields: Awaiting surprises for tests of gravity in space}",
    eprint = "astro-ph/0309300",
    archivePrefix = "arXiv",
    doi = "10.1103/PhysRevLett.93.171104",
    journal = "Phys. Rev. Lett.",
    volume = "93",
    pages = "171104",
    year = "2004"
}

@article{Masso:2005ym,
    author = "Masso, Eduard and Redondo, Javier",
    title = "{Evading astrophysical constraints on axion-like particles}",
    eprint = "hep-ph/0504202",
    archivePrefix = "arXiv",
    reportNumber = "UAB-FT-579",
    doi = "10.1088/1475-7516/2005/09/015",
    journal = "JCAP",
    volume = "09",
    pages = "015",
    year = "2005"
}

@article{Masso:2006gc,
    author = "Masso, Eduard and Redondo, Javier",
    title = "{Compatibility of CAST search with axion-like interpretation of PVLAS results}",
    eprint = "hep-ph/0606163",
    archivePrefix = "arXiv",
    reportNumber = "UAB-FT-605",
    doi = "10.1103/PhysRevLett.97.151802",
    journal = "Phys. Rev. Lett.",
    volume = "97",
    pages = "151802",
    year = "2006"
}

@article{Jaeckel:2006xm,
    author = "Jaeckel, Joerg and Masso, Eduard and Redondo, Javier and Ringwald, Andreas and Takahashi, Fuminobu",
    title = "{The Need for purely laboratory-based axion-like particle searches}",
    eprint = "hep-ph/0610203",
    archivePrefix = "arXiv",
    reportNumber = "DCPT-06-136, DESY-06-188, IPPP-06-68, UAB-FT-612",
    doi = "10.1103/PhysRevD.75.013004",
    journal = "Phys. Rev. D",
    volume = "75",
    pages = "013004",
    year = "2007"
}

@article{Kim:2007wj,
    author = "Kim, Jihn E.",
    title = "{PVLAS experiment, star cooling and BBN constraints: Possible interpretation with temperature dependent gauge symmetry breaking}",
    eprint = "0704.3310",
    archivePrefix = "arXiv",
    primaryClass = "hep-ph",
    reportNumber = "SNUTP-07-005",
    doi = "10.1103/PhysRevD.76.051701",
    journal = "Phys. Rev. D",
    volume = "76",
    pages = "051701",
    year = "2007"
}

\end{document}